\documentclass[sigconf,nonacm,10pt]{acmart}
\AtBeginDocument{%
  }

\usepackage{xspace}
\usepackage{pifont}
\usepackage{array}
\usepackage{subcaption}
\usepackage{amsthm}
\usepackage{etoolbox}
\usepackage{listings}
\usepackage{multirow}
\usepackage{makecell}
\usepackage{enumitem}
\usepackage[most]{tcolorbox}

\setlist{nolistsep}

\newcommand{\para}[1]{\smallskip\noindent\textbf{#1}}

\def\Snospace~{\S{}}

\newcommand{\cmark}{\ding{51}}%
\newcommand{\xmark}{\ding{55}}%

\newcommand{\sys}{Ossim\xspace}

\definecolor{googleRed}{HTML}{EA4335}
\definecolor{googleGreen}{HTML}{34A853}
\definecolor{googleBlue}{HTML}{4285F4}

\begin{document}

\title{\sys: OS-Driven Live Simulation for Cluster-Scale Full-Stack Evaluation}

\author{Yiliang Wan\textsuperscript{1},
Haifeng Sun\textsuperscript{1*},
Yihan Yang\textsuperscript{1},
Jonas Kaufmann\textsuperscript{2}, \\
Antoine Kaufmann\textsuperscript{2},
Jialin Li\textsuperscript{1}}
\affiliation{%
  \vspace{8pt}
  \institution{\textsuperscript{1}National University of Singapore \qquad \textsuperscript{2}Max Planck Institute for Software Systems}
  \country{ }
}
\renewcommand{\authors}{Yiliang Wan, Haifeng Sun, Yihan Yang, Jonas Kaufmann, Antoine Kaufmann, Jialin Li}
\makeatletter
\g@addto@macro\@authornotes{\stepcounter{footnote}\footnotetext{Corresponding author.}}
\makeatother

\renewcommand{\shortauthors}{Wan et al.}

\begin{abstract}

Cluster-scale full-stack simulation is essential for evaluating distributed software stacks and emerging hardware components before deployment.
Such simulation must achieve both full-stack fidelity for the unmodified production stack and the simulation performance required for iterative configuration exploration.
However, no existing method achieves both.
We present \sys, an OS-level approach to cluster-scale full-stack simulation built on top of the Linux virtualization stack.
\sys comprises four subsystems: simulation-oriented scheduling, 
live memory hierarchy management, simulation-aware IPC,
and distributed simulation orchestration.
Together, they coordinate live and modeled components under shared simulated time while controlling interference among co-located live hosts.
These mechanisms point toward \emph{simulation-native OS support}, where simulation control and orchestration become core OS responsibilities.

\end{abstract}



\maketitle

\section{Introduction}

Cluster-scale full-stack simulation is essential for end-to-end evaluation of distributed systems and emerging hardware components before deployment.
Cluster operators rely on it to choose deployment configurations for production workloads such as cloud platforms~\cite{openstack,k8s} and big-data frameworks~\cite{spark,hadoop}.
System designers rely on it to evaluate emerging accelerators~\cite{tpu,accelnet,accelflow}, memory systems~\cite{cxl-real-devices}, and interconnects~\cite{cxl,ualink,nvlink,uec} before they are available at the target scale.

A simulator that supports these use cases must satisfy two properties at once.
The first is \emph{full-stack fidelity}: the simulator runs the production stack without source modification and preserves the cross-layer behavior it exhibits across the OS, runtime, application, distributed coordination layers, and underlying hardware.
The second is \emph{simulation performance}: the simulator completes in a timely fashion to support iterative configuration exploration.

Existing simulation methods satisfy some but not all of the properties required for cluster-scale full-stack simulation.
Discrete-event simulators~\cite{gem5,simics,ns3,omnetpp} provide high fidelity within a single simulation component, but do not run the unmodified distributed software stack across cluster nodes.
Modular composition frameworks~\cite{simbricks,splitsim} recover fidelity by stitching simulators together, but their simulation performance is limited by their slowest components when running complex workloads.
Live simulation shows that native execution under simulated time can provide high simulation performance while preserving fidelity, but existing systems~\cite{phantora,nex} remain workload-specific and do not run the full unmodified end-host stack.

We identify that the high performance of live simulation and the full-stack fidelity of modular composition together suggest a path toward cluster-scale full-stack simulation. 
Our key observation is that the Linux virtualization stack serves as a promising
live-execution foundation for this approach.
In particular, KVM~\cite{kvm}, QEMU~\cite{qemu}, and libvirt~\cite{libvirt} provide functional emulation of the unmodified stack at near-native speed.
However, the stack lacks four capabilities that large-scale live simulation requires:
(1) control over when live components advance under simulated time, 
(2) performance isolation across live components on shared hardware,
(3) communication between live and modeled components, and 
(4) coordination among multiple physical hosts for large-scale simulation.
All these capabilities sit below the user-space boundary and therefore call for OS-level mechanisms.

To this end, we present \sys\footnote{\sys is open source at \href{https://github.com/ossim-project/ossim}{https://github.com/ossim-project/ossim}.}, a live simulator for cluster-scale full-stack evaluation
that combines full-stack fidelity with high simulation performance.
\sys adds OS-level live-simulation support to the Linux virtualization stack through four subsystems.
\emph{Simulation-oriented scheduling} synchronizes live and modeled components under shared virtual time.
\emph{Live memory hierarchy management} limits interference among co-located live components for performance fidelity.
\emph{Simulation-aware IPC} controls cross-component events under virtual time.
Finally, \emph{distributed simulation orchestration} composes per-host mechanisms across machines for large-scale simulation.
Together, the four subsystems define a live-simulation substrate spanning servers in a distributed cluster.

Our preliminary prototype runs the TPC-C benchmark in 90.4 seconds, 
whereas a comparable modular gem5-based setup did not finish within a week.
Across representative workloads, \sys reproduces physical behavior with
promising accuracy.
This demonstrates \sys's feasibility for cluster-scale full-stack simulation.
Beyond this setting, \sys points toward \emph{simulation-native OS support}, where simulation control and orchestration become central OS abstractions rather than user-space services above a simulation-agnostic kernel.

\section{Background and Motivation}
\label{sec:background}

\subsection{Cluster-Scale Full-Stack Simulation}
\label{subsec:requirement}

\emph{Cluster-scale full-stack simulation} is essential before deployment
because building and repeatedly reconfiguring large physical testbeds is costly
and often impractical.
By observing end-to-end behavior across software, hardware, and workload
interactions, it provides a lightweight way to guide both near-term deployment tuning and
longer-term hardware planning.
This helps operators revisit server scale, network layout, and software
configuration as workloads evolve.
Designers can use it to evaluate future components, such as accelerators and
memory tiers, within a full distributed system before such components are
available at scale.

Compared with conventional domain-specific simulators
(e.g., network simulators~\cite{ns3,omnetpp}), 
a cluster-scale full-stack simulator must satisfy two properties.

\para{R1. Full-stack fidelity.}
The simulator must faithfully reproduce the software--hardware interactions
that determine cluster-scale behavior.
We decompose this requirement into two aspects: \emph{full-stack coverage} and
\emph{compatibility}.
\begin{itemize}[leftmargin=*]
\item
Full-stack coverage requires capturing interactions among
the OS, runtime, application, distributed coordination logic, hardware resources, 
and deployment configuration.
These interactions determine cluster-level behavior, so modeling one
layer in isolation or replaying recorded traces can miss the effects that shape
end-to-end performance.
\item
Compatibility requires the simulator to run the production stack as
deployed, without source modification.
This property is important for fidelity since modifying or porting the stack
can introduce artifacts that change the behavior being evaluated.
It also allows production binaries and configurations to be evaluated directly
as the stack evolves.
\end{itemize}

\para{R2. Simulation performance.}
The simulator must run fast enough to support iterative exploration.
Because the configuration space of cluster-scale systems is large and 
workload behavior changes over time, operators cannot rely on a single
decisive run.
Instead, they must evaluate a candidate configuration, observe its end-to-end
behavior, refine the configuration, and repeat.
Therefore, simulated-time progress must proceed at a rate compatible with
end-to-end workload duration.
Otherwise, if one minute of simulated workload takes days to complete, 
the simulator becomes impractical for interactive exploration.

\subsection{Related Simulation Methods}
\label{sec:background:existing}

\para{Discrete-event simulation (DES).}
DES models the target system as a sequence of events processed in simulated time.
Different tools in this family target different layers.
Host-side simulators such as gem5~\cite{gem5} and Simics~\cite{simics} target
cycle-level microarchitectures and provide full software compatibility.
Network-side simulators such as ns-3~\cite{ns3} and OMNeT++~\cite{omnetpp} 
target protocols and topologies.
These tools are mature and widely adopted, and they achieve high fidelity for 
studies within their target layer.

However, their layer specialization prevents them from satisfying the two
requirements in \S\ref{subsec:requirement}.
First, they lack full-stack fidelity.
Host-side simulators miss end-to-end
network behavior, while network-side simulators replace the OS, runtime, and
application layers with traffic generators.
Second, they sacrifice simulation performance since high-fidelity DES usually
requires fine-grained event processing.
For example, simulating one second of a 1,024-server fat-tree DCTCP workload in
ns-3 can take over ten hours~\cite{unison}.

\para{Modular composition.}
Modular frameworks~\cite{simbricks,splitsim} 
address the full-stack fidelity limitation 
by composing independent 
host, switch, accelerator, and other component simulators into a unified multi-layer setup.
By coordinating events and time across components, they recover full-stack fidelity that no single simulator provides.
However, these frameworks still rely mainly on DES-based components, and 
thus can only run as fast as the slowest simulator used.

\para{Live simulation.}
Live simulation addresses the simulation performance limitation
by running parts of the system on real hardware under shared simulated time.
Live components contribute real execution behavior at native speed, and modeled components are reserved for parts that are intentionally varied or unavailable.
Phantora~\cite{phantora} achieves fast and accurate simulation for GPU training clusters.
NEX~\cite{nex} targets broader hardware-accelerated systems, but its demonstrated use cases remain limited to a single server and lack full software-stack compatibility.

\vspace{10pt}
\begin{tcolorbox}[
    enhanced,
    width=\linewidth,
    colback=white,
    colframe=googleRed!25,
    boxrule=1.5pt,
    sharp corners,
    left=8pt,
    right=8pt,
    top=16pt,
    bottom=8pt,
    overlay={
        \node[
            anchor=west,
            fill=googleRed!25,
            inner xsep=8pt,
            inner ysep=3pt,
            font=\normalsize
        ] at ([xshift=12pt]frame.north west) {\textit{\textbf{Opportunity}}};
    }
]
\textit{The high performance of live simulation and the full-stack fidelity of
modular composition together pave the way for cluster-scale full-stack
simulation.}
\end{tcolorbox}

\begin{table}[t]
\centering
\caption{Comparison of simulation methods for cluster-scale full-stack simulation.}
\label{tab:comparison}
\footnotesize
\setlength{\tabcolsep}{2.5pt}
\begin{tabular*}{\columnwidth}{@{\extracolsep{\fill}}llccc@{}}
\hline
Method        & System             & Sim.        & End-host   & Cluster- \\
type          &                    & performance & full-stack & scale    \\
\hline
\multirow{2}{*}{DES}
              & gem5/Simics        & \xmark     & \cmark     & \xmark \\
              & ns-3/OMNeT++       & \cmark     & \xmark     & \cmark \\
\hline
Modular comp. & \makecell[l]{SimBricks/\\SplitSim} & * & \cmark & \cmark \\
\hline
\multirow{3}{*}{Live sim.}
              & Phantora           & \cmark     & $\dagger$  & \cmark \\
              & NEX                & \cmark     & \xmark     & \xmark \\
              & \textbf{\sys}      & \cmark     & \cmark     & \cmark \\
\hline
\multicolumn{5}{@{}l}{\scriptsize * Depends on component simulators. $\dagger$ Runs unmodified ML apps without an OS.}
\end{tabular*}
\end{table}

\section{\sys Design}
\label{sec:overview}

\subsection{\sys Overview}
\label{subsec:challenges}

\para{Key idea: Linux virtualization stack as a live execution foundation.}
The opportunity identified in \S\ref{sec:background:existing} suggests that
cluster-scale full-stack simulation can be achieved by combining the high performance of live simulation with the full-stack fidelity of modular composition.
The missing piece is \emph{a live execution foundation that can run the unmodified
production stack correctly and at native speed}.
Our key observation is that the Linux virtualization stack already provides the
main building blocks for this foundation.

The Linux virtualization stack matches this need through a division of
responsibilities across execution, compatibility, and management.
First, KVM~\cite{kvm} enables \textit{native guest execution} through hardware 
virtualization (e.g., Intel VT-x~\cite{intel64}, AMD-V~\cite{amd64}), 
with each vCPU mapped to a kernel thread.
Second, QEMU~\cite{qemu} provides \textit{device compatibility} in user space, 
including device emulation, VM lifecycle management, and I/O processing through 
dedicated threads.
Together, they allow an unmodified guest OS to run at near-native speed.
Third, libvirt~\cite{libvirt} provides APIs for VM lifecycle management and orchestration.

\para{Challenges.}
However, this stack alone does not provide the control needed for live
simulation.
It offers \emph{functional emulation}, where real software executes correctly,
but execution is still governed by the host wall clock and the host's available
performance.
Turning this stack into a live execution foundation for cluster-scale simulation 
requires addressing four challenges:

\begin{itemize}[leftmargin=*]
\item
\textbf{C1}. \emph{Virtual-time-based coordination}:
advance live execution under shared simulated time.
\item
\textbf{C2}. \emph{Live-component isolation}:
bound interference among co-located live components on the shared host.
\item
\textbf{C3}. \emph{Framework-level coordination}:
allow live and modeled components to interact with each other.
\item
\textbf{C4}. \emph{Scalable orchestration}:
coordinate multiple physical hosts to support large-scale simulation.
\end{itemize}

\para{OS-level simulation control.}
These challenges require OS-level support because their control points lie
below the user-space boundary of existing modular composition frameworks.
For \textbf{C1}, KVM vCPU execution and guest-visible clocks are kernel-managed.
Thus, precise control over live execution cannot be provided by user-space
orchestration and must reside inside the kernel.
For \textbf{C2}, performance isolation depends on privileged host resources,
including CPUs, caches, memory bandwidth, NUMA placement, and interrupts.
As these resources are exposed through kernel interfaces, isolation requires
kernel mechanisms.
For \textbf{C3}, cross-component events become visible to live
guests through virtual-device state updates or injected virtual interrupts.
The timing of this visibility is determined by the kernel.
Therefore, \sys realizes these capabilities as OS-level mechanisms rather than
as a purely user-space composition layer.

\begin{figure}[tp]
\centering
\includegraphics[trim={5 0 5 0},clip,width=.9\linewidth]{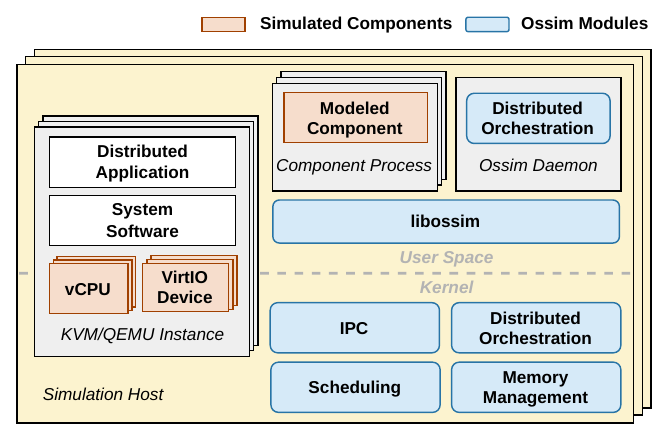}
\caption{The overall architecture of \sys.}
\label{fig:architecture}
\end{figure}

\para{Architecture overview.}
\autoref{fig:architecture} shows the overall architecture of \sys.
\sys follows one principle: 
``keep execution live and make the OS
simulation-aware''.
Live components, i.e., KVM/QEMU instances, execute directly on physical hardware.
Modeled components, i.e., simulated devices, advance according to their
performance models.
\sys coordinates both types of components through four subsystems:
\emph{simulation-oriented scheduling} for virtual-time coordination, 
\emph{live memory hierarchy management} for performance isolation, 
\emph{simulation-aware IPC} (interprocess communication) for
visibility-controlled communication, and 
\emph{distributed orchestration} for
multi-host scale-out.
This allows \sys to preserve full-stack fidelity while avoiding
the fine-grained event processing overhead of pure DES-based simulation.
\autoref{tab:comparison} compares \sys with existing methods.

\subsection{Simulation-Oriented Scheduling}
\label{subsec:scheduling}

\sys uses simulation-oriented scheduling to coordinate live and modeled
components under bounded virtual-time skew (\textbf{C1 addressed}).
The scheduler is built around the \emph{vtask} abstraction, which represents any user-space thread whose execution must be coordinated by the simulation.
Using this unified abstraction, the scheduler \emph{accounts} for vtime
and \emph{dispatches} vtasks only when their progress remains within a
configured skew bound.

\para{Virtual-time accounting.}
For each vtask, \sys maintains a \emph{virtual time} (\textit{vtime}) 
representing its cumulative simulated-time progress.
For live vtasks, vtime is derived from hardware execution,
while for modeled vtasks it is derived from simulated latency
reported by the component model.

\emph{Clock-derived vtime.}
For live vtasks, \sys derives vtime by adapting KVM's \emph{paravirtual clock} (pvclock)~\cite{pvclock}.
In standard KVM, preemption gaps are exposed to the guest through steal time.
\sys instead absorbs these gaps into the hardware TSC (Time Stamp Counter)~\cite{intel64} offset so that
pvclock advances only during actual vCPU execution.
The scheduler reads the same adapted pvclock as the guest, 
providing a single source of truth for both scheduler-maintained vtime and guest-visible time.

\emph{Model-driven vtime.}
For modeled vtasks, \sys lets components report vtime advances computed by
their performance models through either a synchronous \texttt{ioctl} or a
shared per-vtask \emph{run page} for the common asynchronous case.
Modeled components report accumulated simulated latency at configurable granularity, 
and the scheduler advances vtime accordingly. 
\sys also preempts modeled vtasks that
run too long without reporting progress, preventing faulty components from
stalling the simulation.

\para{Dispatch.}
\sys coordinates vtasks through configurable \emph{synchronization domains} (hereafter, domains),
each of which groups vtasks that should progress together within a bounded virtual-time skew.
A vtask may participate in multiple domains.

\emph{Common-case dispatch rule.}
Each domain maintains a cached vtime, defined as the minimum
vtime among its runnable members.
During dispatch, a runnable vtask may execute only if
$\textit{vtask.vtime} \leq \textit{domain.vtime} + \textit{skew\_bound}$.
For vtasks belonging to multiple domains, the condition must hold for every domain.
Vtasks that exceed the skew bound are rescheduled until slower components catch up.
If no vtask on a CPU is eligible, the scheduler yields until other vtasks
advance the relevant domain vtime.
The skew bound controls the trade-off between synchronization precision and
parallelism,
and different domains may use different skew bounds.

\emph{Handling blocked vtasks.}
A blocked vtask is unable to make progress from the scheduler's perspective, 
e.g., while waiting for I/O or halted, and is therefore excluded from the domain minimum.
Including blocked vtasks would artificially pin \textit{domain.vtime} and may deadlock the
simulation, e.g., when halted vCPUs lag behind a running bootstrap vCPU
during VM boot.
When a blocked vtask becomes runnable again, \sys forwards its vtime to the
current domain vtime before allowing it to rejoin bounded-skew scheduling.
This preserves time causality: a vtask that blocks and later wakes observes that
simulated time has advanced, just as a halted physical CPU observes elapsed time
after resuming.

\begin{figure}[t]
\centering
\includegraphics[trim={0 0 0 0},clip,width=1\linewidth]{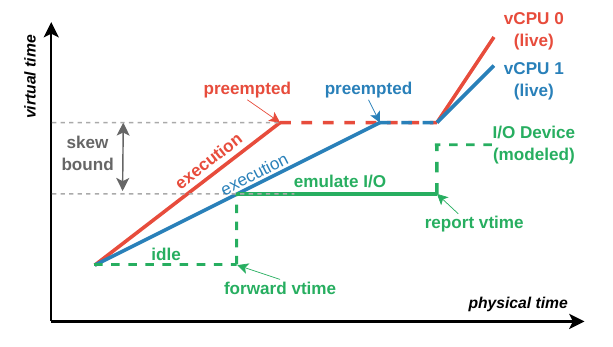}
\caption{An example \sys scheduling timeline.}
\label{fig:scheduling-timeline}
\end{figure}

\autoref{fig:scheduling-timeline} traces these mechanisms within a domain containing two live vCPUs and a modeled I/O device.
The device starts idle and is excluded from the domain minimum.
On wake-up, its vtime is forwarded to the domain minimum, which then preempts the two vCPUs at the skew bound.

\subsection{Live Memory Hierarchy Management}
\label{subsec:isolation}

\sys prevents performance interference among live components through memory
hierarchy management (\textbf{C2 addressed}).
Live KVM hosts execute on real CPUs, so their performance is
sensitive to shared cache capacity, memory bandwidth, and NUMA locality.
By controlling these memory-hierarchy resources, \sys preserves
the fidelity of live execution.

\para{Challenges.}
The subsystem must reduce two kinds of distortion.
The first is \emph{spatial interference}. 
Multiple live hosts running concurrently on the same physical machine
compete for shared caches, memory bandwidth, and NUMA links, 
which slows their execution and therefore corrupts the
virtual-time progress derived from live execution.
The second is \emph{temporal residue}. 
When live hosts time-share the same
hardware, an incoming host may inherit cache and memory state left by a
previously resident host.

\para{Solution.}
\sys organizes live-component isolation around the \emph{cell} abstraction.
A cell is a simulation-oriented generalization of Linux cpusets~\cite{cpusets} that binds one
live host to a controlled resource domain.
This domain includes CPU and NUMA placement, cache capacity,
memory-bandwidth limits, interrupt placement, and cell-switch state management.
For spatial isolation, concurrently active live hosts are assigned to distinct cells.
Cache capacity is controlled via Intel CAT~\cite{intel64} and AMD QoS~\cite{amd64},
memory bandwidth is regulated using MBA~\cite{mba},
and NUMA locality is enforced through NUMA pinning.
For temporal reconditioning, \sys reconditions cache and memory state at cell-switch boundaries by invalidating the outgoing cell's cache footprint and prefetching the incoming cell's working set.
Residual deviations are estimated through PMU sampling and reflected by
adjusting the live CPU's virtual-time advance.
In this way, imperfect isolation is not hidden; it is explicitly incorporated
into simulated time.

\subsection{Simulation-Aware IPC}
\label{subsec:ipc}

The simulation-aware IPC subsystem controls when cross-component messages become
visible, so live and modeled components communicate in simulated-time order
rather than host wall-clock order (\textbf{C3 addressed}).

\para{Challenges.}
The first challenge is preserving communication semantics across heterogeneous
components.
A sender may produce an event at one vtime while the receiver is ahead,
behind, or blocked.
Immediate delivery can therefore expose state at the wrong simulated time and
break causality.
The second challenge is balancing generality with performance.
The IPC abstraction should support simple links, lightweight latency models,
and detailed modeled components, while keeping the common path fast.

\para{Solution.}
\sys organizes simulation-aware IPC around \emph{messages}, \emph{endpoints},
and \emph{hubs}.
Each message separates timing control from data movement.
Its metadata records addressing and virtual-time information, while its payload
is carried directly or through shared memory for high-bandwidth transfer.
Endpoints act as proxies for existing component communication interfaces,
including QEMU device interfaces for live guests.
Hubs reside in the kernel, connecting endpoints and handling lightweight
routing and latency control on the common path.
Their behavior is extensible through eBPF hook points, allowing simple connection
logic to execute inside the hub without expensive context switches.
More detailed connection behavior can instead be modeled as a separate component
through the same endpoint--hub interface with higher overhead.
For each vtask, \sys maintains an incoming queue ordered by visibility time, so that messages become visible in virtual-time order.
The scheduler can use the incoming queue head as a dispatch hint.

\subsection{Distributed Simulation Orchestration}
\label{subsec:orchestration}

The distributed orchestration subsystem scales \sys by 
composing per-host subsystems (virtual-time
coordination, IPC, isolation) into a single
cluster-scale simulation substrate while preserving a unified simulation
view (\textbf{C4 addressed}).

\para{Challenges.}
This subsystem must address two challenges.
First, it must preserve local subsystem semantics across hosts, 
e.g., synchronization domains may contain remote vtasks, and
IPC hubs may connect endpoints on different machines.
Second, it must limit coordination overhead, since cross-host vtime
synchronization and distributed IPC can place network communication on the
simulation hot path.

\para{Solution.}
\sys combines a per-host user-space daemon with kernel-level support
to extend local coordination mechanisms across physical hosts.
To preserve cross-host semantics, remote vtasks are represented as local proxy
vtasks that participate in bounded-skew enforcement, while logical hubs are
implemented as distributed hub instances that exchange messages carrying
addressing and visibility-time metadata.
To reduce coordination overhead, the daemon handles only control-plane tasks,
such as component placement and channel setup, while kernel mechanisms keep
virtual-time updates and IPC delivery on the hot path.
\sys further reduces cross-host traffic by co-locating frequently interacting components when resources permit.

\section{Preliminary Results}
\label{sec:evaluation}

We implement a prototype of \sys and conduct preliminary experiments to evaluate its accuracy and efficiency when simulating unmodified software stacks.

\para{Implementation.}
Our prototype implements the core synchronization mechanism and adds an Ethernet switch hub.
For synchronization, we integrate a customized eBPF scheduler~\cite{schedext} with QEMU/KVM so that vCPU threads across instances advance under shared virtual time.
For inter-component communication, we build the Ethernet switch hub on a Linux bridge with an interface to configure per-link performance characteristics, including bandwidth and latency.
We used coding agents to assist with the implementation and debugging of \sys.

\para{Experiments.}
We connect up to three QEMU/KVM VMs on one host in a star topology via the prototype hub.
As ground truth, we deploy a matching physical testbed connected through a hardware Ethernet switch, with prototype hub parameters set to match.
We evaluate \sys against the physical testbed across three workload categories: a single-host CPU benchmark (CoreMark), database workloads (TPC-C on MySQL, YCSB on HBase), and big-data frameworks (TPC-DS 99 on Spark and on Hive).

\begin{table}[t]
\centering
\caption{Accuracy and wall time on example workloads.}
\label{tab:exp-results}
\footnotesize
\setlength{\tabcolsep}{2.5pt}
\renewcommand{\arraystretch}{1.1}
\begin{tabular*}{\columnwidth}{@{\extracolsep{\fill}}lccccc@{}}
\hline
\multirow{2}{*}{Workload}      & \multirow{2}{*}{\#Inst.} & \multirow{2}{*}{Metric}  & \multirow{2}{*}{Acc.}  & \multicolumn{2}{c}{Wall time (s)}                       \\
\cmidrule(lr){5-6}
                               &                          &                        &          & Real                   & \sys                             \\
\hline
CoreMark                       & 1                        & Arith. Speed           & 97.8\%   & 12.5                   & 14.2 (1.1$\times$)               \\
\multirow{2}{*}{TPC-C (MySQL)} & \multirow{2}{*}{2}       & Avg. Latency           & 82.2\%   & \multirow{2}{*}{60.3}  & \multirow{2}{*}{90.4 (1.5$\times$)} \\
                               &                          & Throughput         & 92.3\% &                  &                                     \\
YCSB (HBase)                   & 3                        & Runtime                & 81.5\%   & 28.7                   & 81.5 (2.8$\times$)               \\
TPC-DS 99 (Spark)              & 3                        & Runtime                & 72.7\%   & 9.3                    & 36.8 (4.0$\times$)               \\
TPC-DS 99 (Hive)               & 3                        & Runtime                & 92.0\%   & 31.8                   & 98.8 (3.1$\times$)               \\

\hline
\end{tabular*}
\end{table}

\para{Results.}
For each workload, \autoref{tab:exp-results} reports \sys's accuracy relative to the physical testbed together with wall-clock execution time.
\sys completes all workloads with a modest slowdown relative to the physical run, ranging from 1.1$\times$ (CoreMark) to 4.0$\times$ (TPC-DS on Spark).
The slowdown increases with the number of synchronized KVM instances
due to additional cross-instance coordination overhead.
For comparison, we also attempted the TPC-C (MySQL) workload with modular simulation using gem5~\cite{gem5} hosts, but the experiment did not finish within a week, whereas \sys completed the same workload in 90.4~seconds.
The prototype achieves up to 97.8\% accuracy relative to the physical metric.
We read this as an encouraging preliminary signal, given that the prototype implements only part of the full design.
Several sources of inaccuracy remain uncontrolled, such as memory-hierarchy interference and IPC visibility timing.
Overall, these preliminary results suggest that \sys can simulate unmodified distributed software stacks with promising accuracy and substantially better performance than existing modular simulation approaches.

\section{Related Work and Discussion}
\label{sec:related}

\noindent\textbf{Simulation acceleration.}
Apart from live simulation, prior work accelerates simulation through system and runtime optimizations~\cite{dons,unison,klonet,legosim,splitsim,ash}, learned performance models~\cite{routenet,routenet-fermi,mimicnet,deepqueuenet,m3}, and specialized hardware such as GPUs and FPGAs~\cite{cunetsim,nsx,gedes,gem,miniature,firesim}.

\para{Simulation for distributed stacks.}
Other work targets specific cluster workloads, particularly GPU training~\cite{simai,spme,phantora}, simulating both the distributed training strategies and the underlying hardware behavior.
In contrast, \sys focuses on unmodified full-stack distributed software and its interaction with underlying hardware at cluster scale.

\para{Resource isolation for guests.}
Partitioning hypervisors~\cite{lpar,nohype,jailhouse,bao} statically bundle fixed host resources for each guest and provide isolation through strict non-sharing of cache and memory hardware.
In contrast, \sys allocates host resources dynamically as simulated guests come and go, and its goal is simulation accuracy under controlled sharing.

\para{Discussion.}
Our current prototype remains preliminary, with several subsystems still under active development. 
Beyond KVM-based VMs, integrating other virtualization technologies~\cite{nvidia-vgpu,xsched,cricket,gpreempt,krypton,gshare,nixie} 
could extend \sys to additional live hardware such as GPUs.

\section{Conclusion}
\label{sec:conclusion}

\sys takes an OS-level approach to cluster-scale full-stack simulation of unmodified distributed software stacks, integrating simulation-oriented mechanisms into the Linux virtualization stack.
Our preliminary prototype reproduces physical behavior with promising accuracy and substantially better performance than existing approaches.
We see \sys as an initial step toward \emph{simulation-native OS support}, a broader direction beyond cluster-scale full-stack simulation.

\begin{acks}
We thank the anonymous reviewers for their valuable comments. This work was
supported by the Singapore Ministry of Education, under the Academic Research
Fund Tier 3 grant MOE-MOET32024-0004.
\end{acks}

\bibliographystyle{ACM-Reference-Format}
\bibliography{references}

\end{document}